\documentclass[pss,fleqn,embeddedheads]{w-art}
\usepackage{times}
\usepackage{w-thm}
\usepackage[]{graphicx}
\usepackage[super]{natbib}
\begin{document}
%%    The information for the title page will be placed between
%%    \begin{document} and \maketitle. The order of most entries
%%    is determined by the class file and can not be changed by
%%    rearranging them. The maketitle command follows after the
%%    absract.
%%
%%    Most of the following commands will be completed by the publisher.
%%
\DOIsuffix{theDOIsuffix}
%%
%% issueinfo for header and copyright line
\Volume{XX}
\Issue{1}
\Copyrightissue{01}
\Month{01}
\Year{2004}
%%
%%    First and last pagenumber of the article. If the option
%%    'autolastpage' is set (default) the second argument may be left empty.
\pagespan{1}{}
%%
%%    Dates will be filled in by the publisher. The 'reviseddate' and
%%    'dateposted' (Published online) entry may be left empty.
%\Receiveddate{\sf zzz} \Reviseddate{\sf zzz} \Accepteddate{\sf
%zzz} \Dateposted{\sf zzz}
%%
%%    Give a maximum of six PACS code in numerical order.

%% \pretitle{Editor's Choice}

%% We have a short and a long form for the title. The short form
%% (optional argument) goes into the running head.

\title[Density functionals and half-metallicity in La$_{2/3}$Sr$_{1/3}$MnO$_3$]
{Density functionals and half-metallicity in La$_{2/3}$Sr$_{1/3}$MnO$_3$}

%% Please do not enter footnotes or \inst{}-notes into the optional
%% argument of the author command. The optional argument will go into
%% the header.  If there is only one address the marker \inst{x} may be
%% omitted.

%% Information for the first author.
\author[V. Ferrari]{V. Ferrari \footnote{Corresponding
     author: e-mail: {\sf vpf20@cam.ac.uk}, Phone: +44\,1223\,337466, Fax:
     +41\,1223\,337356}\inst{1}}
\author[J. M. Pruneda ]{J. M. Pruneda \inst{2}}
\author[Emilio Artacho]{Emilio Artacho\inst{3,4}}
%%\footnote{Third author footnote.} may be inserted after the name.

\address[\inst{1}]{Cavendish Laboratory, University of Cambridge, Madingley
Road, Cambridge CB3 0HE, UK}
\address[\inst{2}]{Institut de Ciencia de Materials de Barcelona, 
CSIC Campus U.A.B., 08193 Bellaterra, Barcelona, Spain}
\address[\inst{3}]{Department of Earth Sciences, University of Cambridge, 
Downing Street, Cambridge CB2 3EQ, UK}
\address[\inst{4}]{Donostia International Physics Centre, Universidad 
del Pa\'{\i}s Vasco, 20080 San Sebastian, Spain}

\begin{abstract}
  The electronic structure and equilibrium geometry of 
La$_{2/3}$Sr$_{1/3}$MnO$_3$ are studied theoretically by means
of density functional calculations.
  The doping is treated by introducing holes and a compensating
jellium background.
  The results for the local density approximation (LDA) agree with previous
LDA calculations, with an equilibrium volume 5.3 \% too small and with both 
majority and minority spin states present at the Fermi level for the relaxed 
system.
  The generalised gradient approximation (GGA) offers a qualitatively
improved description of the system, with a more realistic volume, and a 
half-metallic behaviour for the relaxed structure, which enables studies 
needing theoretical relaxation.
%  The description of the doping is further explored for the ideal 
%(001) surface, for which it is compared with calculations with 
%explicit Sr doping.
The ideal MnO$_2$-terminated (001) surface is then described with 
explicit doping.
\end{abstract}

\maketitle                   % Produces the title.

%% If there is not enough space inside the running head
%% for all authors including the title you may provide
%% the leftmark in one of the following three forms:

%% \renewcommand{\leftmark}
%% {First Author: A Short Title}

%% \renewcommand{\leftmark}
%% {First Author and Second Author: A Short Title}

\renewcommand{\leftmark}
{V. Ferrari: Density functionals for La$_{2/3}$Sr$_{1/3}$MnO$_3$}

\section{Introduction}

  Ferromagnetic perovskite manganites attract much attention because of 
interesting properties like the colossal magnetoresistance found in 
some phases \cite{Ramirez} and the high degree of spin polarization of 
the charge carriers in other phases.
  The optimally doped case of La$_{2/3}$Sr$_{1/3}$MnO$_3$ (LSMO) reaches 
complete polarisation at the Fermi level, the so-called half-metallic 
state.~\cite{Park}
  These features make manganites very appealing for spin-polarized 
current injection in new devices for spintronics, 
such as tunnel junctions.
  First-principles calculations could be very helpful in the understanding
and characterisation of such devices. 

  Descriptions of the electronic structure of bulk LaMnO$_3$ based compounds
within density-functional theory have already been 
documented,~\cite{PickettSingh} mainly based on LDA.
  Electrons can be highly correlated in manganites, and it is known 
that the electronic characteristics of some of their phases would demand
a much more explicit consideration of the non-local exchange (Hund's rule
in the $d$-shell of the Mn ions) than what provided by present-day GGAs.
  LSMO, however, is metallic and thus much more amenable 
to a Kohn-Sham description, as already shown by Pickett.~\cite{PickettSingh}

  Two important limitations have been found in the LDA description of LSMO.
  (i) The theoretical volume obtained from LDA is underestimated by 5.3 \%, 
the material losing its half-metallic character in its relaxed 
geometry.~\cite{problemLDA}
  This limitation has been used to argue for the need of a treatment of
correlation effects beyond LDA, such as the LDA+U 
approach,~\cite{Satpathy, Solovyev} even if the point is still 
debated.~\cite{Eriksson}
  The flawed volume in itself  does not prevent realistic studies of the 
bulk of the pure material, since the geometry can be taken directly from 
experiment. 
  It becomes a very important limitation, however, for any
more complex situation (surface, interface, defect), for which the 
first-principles relaxation of the structure is needed before any other 
property can be analysed.
  (ii) The Sr doping is hard to describe, demanding large unit cells
and statistical sampling of configurations, or alternative alloy 
descriptions like the coherent-phase approximation, which enormously 
complicate any study.

  In this work we explore the GGA description of LSMO,
observing that a much better estimate of the geometry is obtained, 
which maintains a half-metallic character.
  It is important to stress that this is not a fundamental result, since 
we do not know whether this character is retained for the right reason.
  This result is important for practical purposes, since it allows
the use of GGA DFT for structural optimisations of LSMO and its
surfaces, interfaces, defects etc.
  We have also tested the performance of a homogeneous approximation
to hole doping, replacing Sr substitutions by an homogeneous background
of compensating charge, as recently tried in other perovskite 
oxides.~\cite{Andrew}  
  Finally, the GGA description is tried on the ideal (001) surface of LSMO.
The homogeneous background would have no sense for this calculation 
since the Sr substitution should be distributed only within the solid and not
in vacuum.  Hence, we choose simple realisations of explicit doping for this 
latter study.

\section{Method}
\label{method}

  We have performed density functional theory (DFT) \cite{DFT,DFT_Kohm&Sham} 
calculations using the SIESTA method \cite{SIESTAmethod}, which is based on
pseudopotentials and numerical localized atomic orbitals as basis 
sets. 
  The calculations were performed using two approximations to DFT, 
namely, the local density approximation \cite{LDA} (LDA) and the
generalized gradient approximation \cite{GGA} (GGA), both in
the spin-polarised version.
  For LDA we use the Ceperley-Alder exchange-correlation 
potential,~\cite{Ceperley-Alder} whereas for the GGA calculations, 
we use the the Perdew-Burke-Ernzerhof (PBE) scheme.~\cite{GGA}

% \section{Pseudopotentials and Basis Set}
% \label{ppt_basis}

  Core electrons were replaced by norm-conserving pseudopotentials using
the scalar-relativistic Troullier-Martins scheme\cite{Troullier-Martins}
in the Kleinman-Bylander factorised form,~\cite{KB} 
with nonlinear core corrections.~\cite{corecorrections}
  La and Mn are delicate species for pseudopotentials.
  The 5$s^{2}$ 5$p^{6}$ 5$d^{0}$ 4$f^{0}$ and 3$s^{2}$3$p^{6}$ 3$d^{5}$ 
4$f^{0}$ configurations were used as reference for La and Mn respectively.  
  Notice that the 5$s^2$ and 5$p^6$ semicore states of La and the 
corresponding 3$s^2$ and 3$p^6$ states of Mn are explicitly included in 
the calculations.
  The core radii for the $s$, $p$, $d$ and $f$ components of the 
pseudopotential are 1.85 a$_{\rm o}$, 2.20 a$_{\rm o}$, 3.10 a$_{\rm o}$, 
1.40 a$_{\rm o}$ for La, with a radius for the core correction ($r_{pc}$) of 
1.50 a$_{\rm o}$. 
  The radii for Mn are 1.40 a$_{\rm o}$, 1.90 a$_{\rm o}$, 1.50 a$_{\rm o}$, 
and 1.90 a$_{\rm o}$ with $r_{pc}$=0.70 a$_{\rm o}$.
  For O and Sr we use the pseudopotentials from previous work.~\cite{Junquera}
  For the GGA O pseudopotential we use a slight partial core correction to 
avoid the GGA instability close to the nucleus.~\cite{Haman}

  The basis set consists of strictly localised atomic orbitals at the 
double-$\zeta$ polarised level.~\cite{Artacho99}
  The semicore states are treated at the single-$\zeta$ level.
  The radial shape of the orbitals is determined variationally\cite{Junquera} 
under a confining pressure of 0.2 GPa,~\cite{Anglada} using cubic bulk 
LaMnO$_3$ as reference for La and Mn, and using a H$_2$O molecule for O.
  Pseudopotentials and the basis were tested by comparing our results 
for cubic LaMnO$_3$ with those in the literature both for LDA and GGA
\cite{PickettSingh,Eriksson}.

  The Sr doping is treated in two ways. 
  Firstly, by adding the desired concentration of holes to the pristine
system, plus a compensating homogeneous negative background.~\cite{Andrew}
  This is justified by the main finding that the doping produces
mainly a change in band filling with little electronic or geometric
distortion otherwise. 
  Secondly, the results for that approximation are compared with 
calculations with explicit Sr doping.

\section{Half-metallicity in bulk LSMO}
\label{halfmetal}

\begin{SCfigure}
%\begin{SCfigure}[4][b]
\includegraphics[width=.53\textwidth]{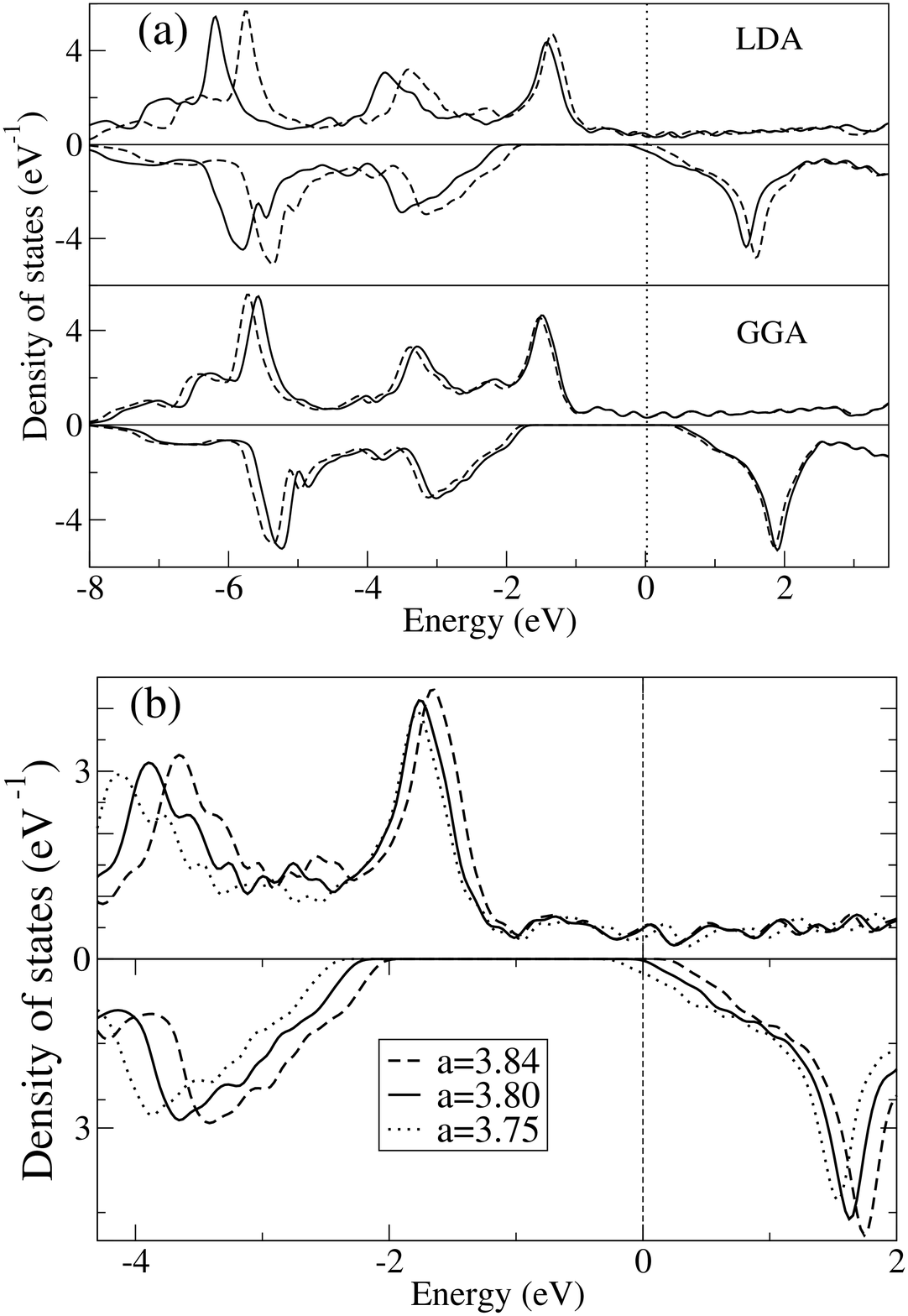}
\caption{(a) Comparison of the density of electronic states for the LDA and
GGA approximations. 
  Majority (minority) spin states are depicted in the upper (lower) graph of
each panel.
  Solid lines correspond to the theoretical equilibrium lattice constant,
while dashed lines are for the pseudo-cubic experimental value.
(b) Density of electron states around the Fermi level for three 
different values of the lattice constant, within the GGA approximation.}
\vspace{1cm}
\label{LDAvsGGA}
\end{SCfigure}

  In this section we consider the effect of the density functional in the 
lattice constant and the half-metallic character of bulk LSMO.
  Although LDA is able to reproduce the half-metallic behaviour for 
the experimental pseudo-cubic lattice parameter \cite{LaSrMnO3experimental}
($a_{exp}=$3.89 \AA), this is not the case when the theoretical 
equilibrium lattice constant is used ($a_{eq}^{LDA}=$3.82 \AA), 
as can be observed in Fig.\ref{LDAvsGGA} (a).  
  The origin of the problem is the underestimation of the lattice
parameter ($\sim$2 \%), for which the minority-spin band gap is smaller.
  Within GGA, however, the half-metallic character is maintained for both the 
experimental and the theoretical lattice constants ($a_{eq}^{GGA}=$3.91 \AA), 
which display a much better agreement (an overestimation of $\sim 0.5 \%$)
and thus allow the use of GGA for structure optimisation.
  The dependence of the half-metallicity on volume is further explored
in Fig.~\ref{LDAvsGGA} (b), where it is shown that for values of the 
lattice constant below $\sim$ 3.80 \AA\ the half-metallic character is lost.
  GGA is used in the calculations presented in the rest of the paper.

\begin{SCfigure}
\includegraphics[width=.53\textwidth]{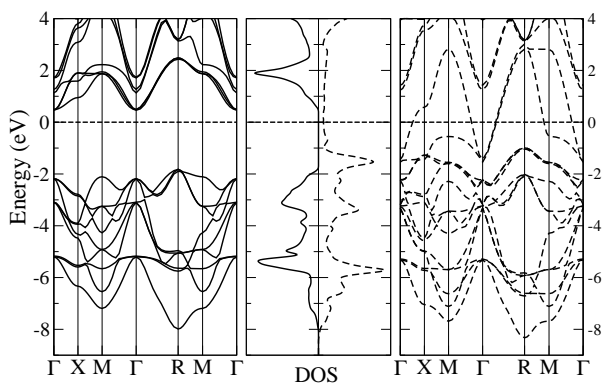}
\caption{Band structure and density of states (centre panel) for hole-doped 
LaSrMnO$_3$. Minority (majority) spin is shown with solid (dashed) lines.} 
\label{bandstructure}
\end{SCfigure}

  The GGA band structure of LSMO is shown in Fig.~\ref{bandstructure}.
  The bands near the Fermi level are mainly due to hybridization of the 
oxygen 2$p$ and the manganese 3$d$ orbitals. 
  The conduction bands at the Fermi level correspond to the 
occupied $e_g$ majority spin states. The peaks in the DOS at 
$\sim 1.5$ eV below $E_F$ and $\sim 2$ eV above $E_F$ correspond to the
$t_{2g}$ states for the majority (occupied) and minority (unoccupied) spins.

%\begin{SCfigure}[4][b]

\section{Surface properties} 
\label{surface}
  For the study of the (001) surface of LSMO we have chosen MnO$_2$ 
termination.~\cite{MnOtermination}
  A slab with 6 unit cells in the perpendicular direction (12 layers 
of MnO$_2$-La(Sr)O planes) and 4x2 unit cells in the parallel 
directions was used to model the system. 
  The chosen thickness ensures that the bulk properties are recovered 
at the center of the slab, and the in-plane repetition was introduced 
for the description of explicit 1/3 doping, by substituting one La
by Sr in every three.

\begin{SCfigure}
\includegraphics[width=.53\textwidth]{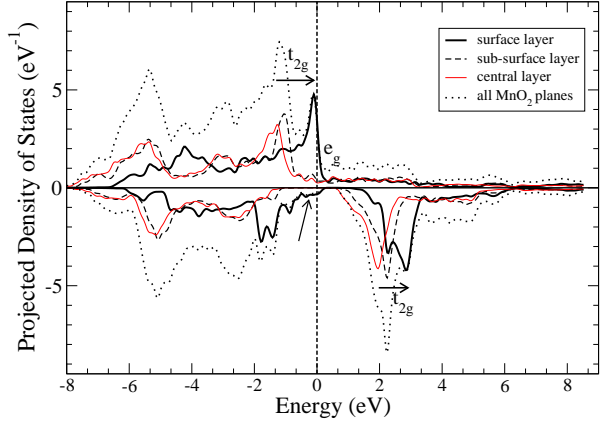}
\caption{Projected Density of States for different MnO$_2$ layers in 
the (001) slab. Majority (minority) spin is shown in the upper (lower) panel. 
  The shifts in the $t_{2g}$ peaks as well as the partial occupation of the
minority spin for the surface layer are highlighted by arrows.} 
\label{MnO_layers}
\end{SCfigure}

  In Fig. \ref{MnO_layers} we plot the projected density of states (PDOS) 
on the MnO$_2$ layers at different depths. 
  The PDOS on a layer is obtained by summing the PDOS for all the basis
orbitals associated to that layer, and
$$PDOS_{\mu}(E)=\sum_{n,\nu} {C^*_{\mu n} C_{\nu n} 
S_{\nu \mu} \delta (E-E_n)} , $$
where $E_n$ and $\psi_n(\vec r)=\sum_{\mu} {C_{\mu n} \phi_{\mu}(\vec r)}$ 
are the Kohn-Sham eigenvalues and eigenvectors, respectively, the latter
expanded on the atomic basis $\{\phi_{\mu}(\vec r)\}$, with overlap
$S_{\nu \mu}$.
  The surface layer introduces a strong peak near the Fermi level mainly 
for the majority spin but with some weight on the minority spin as well
that breaks the half-metallicity of the system.  
  This deviation is very localised close to the surface. 
  For the subsurface MnO$_2$ layer the PDOS already resembles that of the 
bulk and for the central layer the half-metallic character is fully recovered.
  This surface contamination of the spin polarisation was found in other 
calculations of doped manganites and is related to magnetic reconstructions 
at the surface due to changes in the electronic occupation for the 
$d_{z^2}$ and $d_{xy}$ states.~\cite{Filippetti}  
  The $t_{2g}$ states in the surface are shifted upward by about 1 eV 
and overlap with the $e_g$ states at the Fermi level. 
  This changes the magnetic interaction and, although the system remains 
ferromagnetic, the magnetic moments near the surface are slightly reduced 
($\sim$3.4$\mu_B$ as compared to $\sim$3.5$\mu_B$ in bulk). 
  This reduction could be related to the experimental evidence of a
reduction in the magnetization at the surface of LSMO, but is in 
contradiction with similar theoretical calculations for doped 
manganites that predict an increase in the magnetic moment at 
the surface.~\cite{Filippetti}
A full characterisation of the surface state and its dependence with
termination and surface relaxation will be presented
elsewhere.~\cite{Valeria}

\section{Conclusions}

  We have studied the bulk properties of hole-doped 
La$_{2/3}$Sr$_{1/3}$MnO$_3$ comparing LDA and GGA (PBE).
LDA gives a poor description of both the structure (lattice parameter) and
the electronic properties (half-metallicity) of the system.  
GGA reproduces the half-metallic character and gives lattice 
constants in good agreement with experiment.
The fact that the full (variable cell) relaxation of the bulk 
under GGA retains the half-metallic character allows obtaining 
relaxed geometries within GGA for surfaces, interfaces, or 
nanostructures of any sort.

\begin{acknowledgement}
We thank N. Spaldin, P. Littlewood, N. Mathur, L. Hueso and L. Brey 
for very useful discussions.  
EA acknowledges the hospitality at the Donostia International Physics Centre.  
We acknowledge financial support from EPSRC, NERC, and BNFL.
\end{acknowledgement}

%\begin{thebibliography}{10}
%\bibitem[1]{wang}
%\end{thebibliography}

%\begin{thebibliography}{10}
%\bibitem[1]{wang} Q.~Wang, Y.~Yang, and R.~Wang,
%phys.~stat.~sol.~(a) {\bf 155}, 289 (1996).
%\bibitem[2]{deak} P.~De´ak, Th.~Frauenheim,
%and M.~R.~Pederson (eds.), Computer Simulation of Materials at
%Atomic Level (Wiley-VCH, Berlin, 2000), p. 89.
%\bibitem[3]{ibach} H.~Ibach, Electron Energy Loss Spectrometers: The
%Technology of High Performance, Springer Series in Optical
%Sciences Vol.~63 (Springer, Berlin, 1990), chap.~5.
%\bibitem[4]{procbook} B.~J.~Ansell, I.~Harrison, and C.~T.~Foxon, in:
%Proceedings of the 4th International Conference on Nitride
%Semiconductors, Denver, Colorado, USA, 2001, Part~A (Wiley-VCH,
%Berlin, 2002), pp.~279--282.
%\bibitem[5]{procjnl} B.~J.~Ansell et al., Proceedings of the 4th International Conference on Nitride
%Semiconductors, Denver, Colorado, USA, 16--20 July 2001, Part~B.1;
%phys.~stat.~sol.~(a) {\bf 188}, 279 (2001).\\...
%\bibitem[10]{pss03} phys.~stat.~sol.~(b) {\bf 235} ff. (2003);
%phys.~stat.~sol.~(a) {\bf 195} ff. (2003); phys.~stat.~sol.~(c)
%{\bf 0} (2003).
%\end{thebibliography}

\bibliographystyle{apsrev}
\bibliography{article}

\end{document}